\documentclass[prd,draft,showpacs,superscriptaddress,nofootinbib]{revtex4}
\usepackage[T1]{fontenc}
\usepackage{amssymb,amsmath,amsfonts}
\usepackage{mathrsfs,mathtools}
\usepackage{ccaption}
\usepackage{graphicx}
\usepackage{tikz}
\usetikzlibrary{decorations.pathmorphing,calc}

\def\clock{{\count0=\time
           \divide\count0 60
           \ifnum\count0<10 0\fi\the\count0
           \multiply\count0 -60 \advance\count0 \time
           :\ifnum\count0<10 0\fi \the\count0
         }}
\newcommand{\timestamp}{{\small\vbox{\hbox{\tt\jobname.tex}
\hbox{\the\day/\the\month/\the\year, \clock}}}}

\usepackage{amsmath,amssymb}
\usepackage{verbatim}
\usepackage{graphicx}
\usepackage{hyperref}
\usepackage{color}

\DeclareFontFamily{OT1}{rsfs}{}
\DeclareFontShape{OT1}{rsfs}{m}{n}{ <-7> rsfs5 <7-10> rsfs7 <10->rsfs10}{} 
\DeclareMathAlphabet{\mycal}{OT1}{rsfs}{m}{n}

\newcommand{\be}[1]{ \begin{equation}\label{#1} }
\newcommand{\ee}{\end{equation}}
\newcommand{\bea}[1]{\begin{eqnarray}\label{#1} }
\newcommand{\eea}{\end{eqnarray}}

\renewcommand{\phi}{\varphi}
\newcommand{\eq}[2]{\begin{equation} #1 \label{#2} \end{equation}}

\newcommand{\de}{\delta}

\DeclareMathOperator{\extdm}{d}
\newcommand{\extd}{\extdm \!}

\DeclareRobustCommand{\rchi}{{\mathpalette\irchi\relax}}
\newcommand{\irchi}[2]{\raisebox{\depth}{$#1\chi$}}

\begin{document}

\title{Flat Space Limit of (Higher-Spin) Cardy Formula}

\author{Max Riegler}
\email{rieglerm@hep.itp.tuwien.ac.at, max.riegler@yukawa.kyoto-u.ac.jp}
\affiliation{Institute for Theoretical Physics, Vienna University
of Technology, Wiedner Hauptstrasse 8-10/136, A-1040 Vienna,
Austria, Europe}
\affiliation{Yukawa Institute for Theoretical Physics (YITP), Kyoto University, Kyoto 606-8502, Japan}

\date{\today}

\preprint{TUW-14-12}
%%%%%%%%%%%%%%%%%%%%%%%%%%%%%%%%%%%%%%%%%%%%%%%%%%%%%%%%%%%%%%%%%%%%%%%%%%%%%%

\begin{abstract}
In this note I derive the flat space limit of the modified Cardy formula associated with inner horizons and show that it reproduces the correct Galilean conformal field theory counting of flat space cosmology microstates. l also determine the entropy of flat space cosmologies in flat space chiral gravity in this way. In addition, I derive a Cardy-like expression for flat space cosmologies with spin-3 charges and thus give a prediction for the corresponding Galilean conformal field theory counting of flat space cosmology microstates with spin-3 charges.
\end{abstract}

\pacs{04.60.Rt, 04.60.Kz, 11.25.Tq}

\maketitle

\paragraph{Introduction.} One of the most important tools in gaining a better understanding of quantum gravity is the holographic principle \cite{Susskind:1994vu} which states that a quantum theory of gravity in $d+1$ dimensions has an equivalent description in terms of dual quantum field field theory without gravity in $d$ dimensions. The Anti-de Sitter/Conformal Field Theory (AdS/CFT) correspondence which relates a type IIB string theory on AdS$_3\times S^5$ with an $\mathcal{N}=4$ super Yang-Mills theory \cite{Maldacena:1997re} is an explicit realization of this holographic principle. Since the first appearance of this explicit example of holography a lot of research has been done in this field, especially on spacetimes which are AdS.\\
Recently there has also been a lot of interest in establishing a holographic correspondence between $2+1$ dimensional asymptotically flat spacetimes \cite{Barnich:2006av} and $2$ dimensional field theories with symmetries given by the Galilean conformal algebra (GCA) \cite{Bagchi:2009my,Bagchi:2010zz}. An especially interesting observation is that many results that can be obtained via an AdS$_3$/CFT$_2$ correspondence can be used to establish a flat space/Galiliean conformal algebra correspondence by performing an \.In\"on\"u--Wigner contraction i.e.~an ultra-relativistic boost \cite{Bagchi:2012cy,Barnich:2012aw,Krishnan:2013wta}.\\
One can take as an explicit example for this AdS$_3$ with Brown-Henneux boundary conditions \cite{Brown:1986nw} where the symmetry algebra is given by two copies of the Virasoro algebra with generators $\mathfrak{L}_n$ and $\bar{\mathfrak{L}}_n$ and central charges $c$, $\bar{c}$ respectively. Defining the following set of generators and taking the limit $\ell\rightarrow\infty$, where $\ell$ is the AdS radius 
	\begin{align}\label{eq:Contraction}
		L_n&=\mathfrak{L}_n-\bar{\mathfrak{L}}_{-n},& M_n&=\frac{1}{\ell}\left(\mathfrak{L}_n+\bar{\mathfrak{L}}_{-n}\right),\\
		c_{L}&=c-\bar{c},& c_{M}&=\frac{1}{\ell}\left(c+\bar{c}\right),
	\end{align}
one readily obtains the Galilean conformal algebra \cite{Bagchi:2009my,Barnich:2006av}
	\begin{align}\label{eq:BMS3}
 		[L_n,\,L_m]&=(n-m)\,L_{n+m} + \frac{c_{L}}{12}\,\big(n^3-n\big)\,\de_{n+m,\,0}, \\
	 	[L_n,\,M_m]&=(n-m)\,M_{n+m} + \frac{c_{M}}{12}\,\big(n^3-n\big)\,\de_{n+m,\,0} \,.
	\end{align}\\
In the holographic context black hole solutions are of special interest because they correspond to thermal states in the dual field theory \cite{Witten:1998zw}. Using the holographic correspondence it is possible to microscopically compute the black hole entropy from the asymptotic growth of states of the dual CFT \cite{Strominger:1997eq} using the Cardy formula \cite{Cardy:1986ie}. The flat limit of the BTZ black hole also corresponds to a thermal state, whose entropy should then be determined microscopically through some counting of states in the dual Galilean CFT (GCFT).\\
It has been shown \cite{Cornalba:2002fi} that it is indeed possible to take a meaningful flat limit of the BTZ black hole and end up with a spacetime that has a cosmological horizon which is called a flat space cosmology (FSC). The flat limit in this case is taken as follows. Starting with the BTZ black hole
	\begin{equation}
		\extd s^2=-\frac{(r^2-r_+^2)(r^2-r_-^2)}{r^2\ell^2}\extd t^2+\frac{r^2\ell^2}{(r^2-r_+^2)(r^2-r_-^2)}\extd r^2+r^2\left(\extd\phi-\frac{r_+r_-}{\ell r^2}\extd t\right)^2,
	\end{equation}
with $r_\pm=\sqrt{2G\ell(\ell M+J)}\pm\sqrt{2G\ell(\ell M-J)}$, where $G$ is Newton's constant, $M$ and $J$ are the mass and the angular momentum of the BTZ black hole. Now taking the limit $\ell\rightarrow\infty$ the outer horizon scales as $r_+\rightarrow\ell\hat{r}_+$ in this limit, where $\hat{r}_+=\sqrt{8GM}$, and is thus pushed to infinity. The inner horizon survives the limit and takes the value $r_-\rightarrow r_0=\sqrt{\frac{2G}{M}}J$.
The resulting metric
	\begin{equation}
		\extd s^2=\hat{r}_+^2\extd t^2-\frac{r^2\extd r^2}{\hat{r}_+^2(r^2-r_0^2)}+r^2\extd\phi^2-2\hat{r}_+r_0\extd t\extd\phi,
	\end{equation}
is a FSC with a cosmological horizon at $r_0$.\\
As in the BTZ case one can determine the Hawking temperature $T_H$ from surface gravity and the entropy of the cosmological horizon of the FSC, $S_{FSC}$, via the Bekenstein-Hawking area law. One then observes that the charges (mass $M$ and angular momentum $J$) associated with the FSC obey a first law of thermodynamics
	\begin{equation}
		\extd M=-T_H\extd S_{FSC}+\Omega_{FSC}\extd J,
	\end{equation}
where $\Omega_{FSC}$ is the angular velocity of the FSC horizon. The weird sign in front of the $T_H\extd S_{FSC}$ term is a remainder of the fact that the first law of FSC arises as a limit from the BTZ inner horizon dynamics \cite{Castro:2012av,Detournay:2012ug}. This is an indication that the information regarding the entropy of the FSC is encoded in a contraction of the entropy of the \emph{inner} horizon of a BTZ black hole. This will be crucial for my main goal, when taking the limit of the (inner horizon) Cardy formula.\\

\paragraph{\.In\"on\"u--Wigner Contraction of the Inner Horizon Cardy Formula.} Now I proceed with one of the main goals of this note, namely to show that one can also get the entropy for FSCs by making an \.In\"on\"u--Wigner contraction of the \emph{inner horizon} Cardy formula. While the following calculation itself is simple, it is an essential and new point of this note that one has to take the \emph{inner horizon} limit of the Cardy formula and not the well known result for the outer BTZ horizon. I will first show that taking the flat space limit of the (standard) outer horizon Cardy formula does not yield the correct result for the microstate counting of FSCs.
The Cardy formula that determines the entropy of the outer horizon of a BTZ black hole is given by \cite{Cardy:1986ie}:
\eq{
S_{\textrm{\tiny A}_{\textrm{out}}}=\frac{A_{\textrm{\tiny out}}}{4G}=2\pi\sqrt{\frac{c\,\mathcal{L}}{6}}+2\pi\sqrt{\frac{\bar{c}\,\bar{\mathcal{L}}}{6}}=S_{\textrm{\tiny outer}}
}{eq:cl0}
The left hand side is the Bekenstein--Hawking entropy associated with the outer horizon $A_{\textrm{\tiny out}}$, while the right hand side is the Cardy formula for outer horizons.
The central charges $c$ and $\bar{c}$ are given by
	\begin{equation}\label{eq:cl2}
[\mathfrak{L}_n,\,\mathfrak{L}_m]=(n-m)\,\mathfrak{L}_{n+m}+ \frac{c}{12}\,\big(n^3-n\big)\,\de_{n+m,\,0},\quad
[\bar{\mathfrak{L}}_n,\,\bar{\mathfrak{L}}_m]=(n-m)\,\bar{\mathfrak{L}}_{n+m}+ \frac{\bar{c}}{12}\,\big(n^3-n\big)\,\de_{n+m,\,0},
	\end{equation}
where $\mathcal{L},\bar{\mathcal{L}}$ and their respective conformal weights $\mathfrak{h},\bar{\mathfrak{h}}$ when acting on a highest weight state $|\mathfrak{h},\bar{\mathfrak{h}}\rangle$ are related to the mass and angular momentum in the usual way
	\begin{equation}\label{eq:MassAngMomRel}
		\mathcal{L}=\mathfrak{h}=\frac{1}{2}(M\ell-J),\quad \bar{\mathcal{L}}=\bar{\mathfrak{h}}=\frac{1}{2}(M\ell+J).
	\end{equation}
In order to perform the \.In\"on\"u--Wigner contraction I make the same identifications as in \eqref{eq:Contraction} which in turn also means that the eigenvalues, $h_{L}$, $h_{M}$, of $L_0$ and $M_0$  when acting on a highest weight state $|h_{L},h_{M}\rangle$ are given by
\begin{equation}\label{eq:Contractionhh}
 h_{L} = \mathfrak{h}-\bar{\mathfrak{h}},\quad  h_{M} = \frac1\ell\,\big(\mathfrak{h}+\bar{\mathfrak{h}}\big).
\end{equation}
In addition I define the quantities $\mathcal{M}$ and $\mathcal{N}$ in a similar fashion as in \cite{Barnich:2012aw}
\begin{equation}\label{eq:ContractionMN}
\mathcal{M}=12\left(\frac{\mathcal{L}}{c}+\frac{\bar{\mathcal{L}}}{\bar{c}}\right),\quad \mathcal{N}=6\ell\left(\frac{\mathcal{L}}{c}-\frac{\bar{\mathcal{L}}}{\bar{c}}\right).
\end{equation}
Expressing $c,\,\bar{c},\,\mathcal{L}$ and $\bar{\mathcal{L}}$ in terms of GCA quantities and inserting into $S_{\textrm{\tiny outer}}$ from \eqref{eq:cl0} yields
\eq{
S_{\textrm{\tiny outer}} = 2\pi\sqrt{\frac{c\,\mathcal{L}}{6}}+2\pi\sqrt{\frac{\bar{c}\,\bar{\mathcal{L}}}{6}} =\frac{\pi}{6}\ell\,c_M\sqrt{\mathcal{M}}+ {\cal O}(1/\ell)\,,
}{eq:cl3}
which is obviously not the correct result as its $\ell\rightarrow\infty$ limit diverges with $\ell$.\\
The reason why the flat space limit of the (standard) Cardy formula does not yield the correct GCA result is that the outer BTZ horizon, for which this Cardy formula is valid, gets pushed to infinity in the flat space limit. Taking this into account it is easy to understand why \eqref{eq:cl3} diverges in the flat space limit.\\
Since the cosmological horizon of a FSC is obtained as a limit of the inner BTZ horizon one should thus consider a modified Cardy formula for the BTZ. This modified Cardy formula should count the microstates of the inner BTZ horizon in order to be a valid starting point for a flat space contraction. This modified Cardy formula that determines the entropy of the inner horizon of a BTZ black hole is given by \cite{Castro:2012av,Detournay:2012ug}:
\eq{
S_{\textrm{\tiny A}_{\textrm{int}}}=\frac{A_{\textrm{\tiny int}}}{4G}=\left|2\pi\sqrt{\frac{c\,\mathcal{L}}{6}}-2\pi\sqrt{\frac{\bar{c}\,\bar{\mathcal{L}}}{6}}\right|=S_{\textrm{\tiny inner}}
}{eq:cl1}
The modification in comparison to the (standard) Cardy formula \eqref{eq:cl0} consists of a relative minus sign between the right- ($\mathcal{L}$) and left- ($\bar{\mathcal{L}}$) moving contributions.\\
In order to perform the \.In\"on\"u--Wigner contraction I repeat the same steps as before but I use now \eqref{eq:cl1} instead of \eqref{eq:cl0}. This yields
\eq{
S_{\textrm{\tiny inner}} =\left| 2\pi\sqrt{\frac{c\,\mathcal{L}}{6}}-2\pi\sqrt{\frac{\bar{c}\,\bar{\mathcal{L}}}{6}} \right|= \frac{\pi}{6}\left|c_L\sqrt{\mathcal{M}}+c_M\frac{\mathcal{N}}{\sqrt{\mathcal{M}}}\right| + {\cal O}(1/\ell)\,.
}{eq:cl3}
Taking the $\ell\to\infty$ limit gives a prediction for the microscopic entropy in a Galilean conformal field theory:
\eq{\boxed{
S_{\textrm{GCA}} =  \frac{\pi}{6}\left|c_L\sqrt{\mathcal{M}}+c_M\frac{\mathcal{N}}{\sqrt{\mathcal{M}}}\right|=\pi\,\sqrt{\frac{c_{M}h_{M}}{6}}\,\left|\frac{h_{L}}{h_{M}}+\frac{c_{L}}{c_{M}}\right|}
}{eq:cl4}
This is one of the main results of this Note and agrees precisely with the results obtained in \cite{Bagchi:2012xr,Barnich:2012xq,Bagchi:2013qva}.\\
I compare now with the results for Einstein gravity. In this case $c=\bar{c}$ and hence $c_{L}=0$. The expression \eqref{eq:cl4} then simplifies to
\eq{
S_{\textrm{GCA}}^{\textrm{Einstein}} = \frac{\pi}{6}\left|c_M\frac{\mathcal{N}}{\sqrt{\mathcal{M}}}\right|=2\pi\left|h_L\right|\sqrt{\frac{c_M}{24\,h_M}}
}{eq:cl5}
The result \eqref{eq:cl5} (after translating conventions for $c$-normalization) agrees perfectly with the results in \cite{Bagchi:2012xr,Barnich:2012xq}.\\

\paragraph{Flat Space Chiral Gravity.} One can also use the contractions \eqref{eq:Contraction} and \eqref{eq:Contractionhh} to determine the microscopic entropy of flat space chiral gravity (FS$\rchi$G), a theory that can be obtained as a limit \cite{Bagchi:2012yk} of topologically massive gravity (TMG) \cite{Deser:1982vy}:
	\begin{equation}
		I_{TMG}=\frac{1}{16\pi G}\int\extd^3x\sqrt{-g}\left(R+\frac{1}{\mu}CS(\Gamma)\right).
	\end{equation}
$G$ is again Newton's constant in $2+1$ dimensions, $R$ the Ricci scalar, $\mu$ the Chern-Simons coupling and $CS(\Gamma)=\varepsilon^{\lambda\mu\nu}\Gamma^\rho{}_{\lambda\sigma}\left(\partial_\mu\Gamma^\sigma{}_{\rho\nu}+\frac{2}{3}\Gamma^\sigma{}_{\mu\tau}\Gamma^\tau{}_{\nu\rho}\right)$ is the gravitational Chern-Simons term. Flat space chiral gravity arises in the limit $G\rightarrow\infty$ while keeping fixed $\mu G$ so that $\mu G=\frac{c_{L}}{3}$ remains finite. This is particularly interesting as the central charges of the dual field theory are of a form that allow for unitary representations of the GCA \cite{Grumiller:2014lna} i.e. $c_{L}\neq0,\quad c_{M}=0$.
In \cite{Bagchi:2013qva} it has been shown that the entropy formula for FSCs in TMG take exactly the same form as \eqref{eq:cl4} but with $c_{L}=\frac{3}{\mu G}$, $c_{M}=\frac{3}{G}$, $h_{L}=M+\frac{1}{8G}$ and $h_{M}=J+\frac{M}{\mu}$.
In this limit it is easy to see that $c_{M}\rightarrow0$ and $\sqrt{\frac{c_{M}}{h_{M}}}=\sqrt{\frac{c_{L}}{h_{L}}}$.
Thus, the entropy for FSCs in flat space chiral gravity is given by
	\begin{equation}\label{eq:FSCGEntropy}
		\boxed{S_{\textrm{GCA}}^{\textrm{FS}\rchi\textrm{G}} = 2\pi\sqrt{\frac{c_{L} h_{L}}{6}}=S_{\textrm{CFT}}^{\textrm{Chiral}}.}
	\end{equation}
The result \eqref{eq:FSCGEntropy} coincides precisely with what one would expect of one chiral half of a CFT \cite{Bagchi:2013lma}. This fits very nicely with the suggestion that flat space chiral gravity is indeed the chiral half of a CFT \cite{Bagchi:2012yk,Bagchi:2013hja}.\\

\paragraph{\.In\"on\"u--Wigner Contraction of a BTZ Black Hole with Spin-3 Charges.} As a final result of this note I will now also present a method that allows one to make a prediction for a Cardy-like formula for FSCs that carry spin-3 charges \cite{Afshar:2013vka} as a flat space limit of a rotating BTZ black hole with spin-3 charges $\mathcal{W}$ and $\bar{\mathcal{W}}$.\\
Following \cite{Ammon:2011nk} one can write a Cardy-like formula for the \emph{outer} horizon entropy of a spin-3 charged BTZ as
	\begin{equation}\label{eq:SouterBTZSpin3}
		S_{\textrm{\tiny outer}} =2\pi\left(\sqrt{\frac{c\,\mathcal{L}}{6}}\sqrt{1-\frac{3}{4C}}+\sqrt{\frac{\bar{c}\,\bar{\mathcal{L}}}{6}}\sqrt{1-\frac{3}{4\bar{C}}}\right),
	\end{equation}
where $C$ and $\bar{C}$ are dimensionless constants defined via
	\begin{equation}\label {eq:XiMagicFormula}
		\sqrt{\frac{c}{6\mathcal{L}^3}}\frac{\mathcal{W}}{4}=\xi=\frac{C-1}{C^{\frac{3}{2}}},\quad\sqrt{\frac{\bar{c}}{6\bar{\mathcal{L}}^3}}\frac{\bar{\mathcal{W}}}{4}=\Bar{\xi}=\frac{\bar{C}-1}{\bar{C}^{\frac{3}{2}}},
	\end{equation}
where the $C\rightarrow\infty\,(\bar{C}\rightarrow\infty)$ limit corresponds to the limit of vanishing spin-3 charges. In order to successfully perform a contraction that yields a Cardy-like formula for a spin-3 charged FSC one has again to determine the \emph{inner} horizon spin-3 BTZ formula. In addition one has to find an expression of $C$ and $\bar{C}$ in terms of flat space analogues of these constants which I will call $\mathcal{R}$ and $\mathcal{P}$. This is acutally a non-trivial problem since $C$ and $\bar{C}$ are related to the canonical charges ($\mathcal{L},\bar{\mathcal{L}},\mathcal{W},\bar{\mathcal{W}}$) in a non-linear way. In the following I will, however, present a method to solve this problem. First I will introduce the flat space analogues of the spin-3 charges $\mathcal{W}$ and $\bar{\mathcal{W}}$ in analogy to the spin-2 case as
\begin{equation}\label{eq:ContractionVZ}
\mathcal{V}=12\left(\frac{\mathcal{W}}{c}+\frac{\bar{\mathcal{W}}}{\bar{c}}\right),\quad \mathcal{Z}=6\ell\left(\frac{\mathcal{W}}{c}-\frac{\bar{\mathcal{W}}}{\bar{c}}\right).
\end{equation}
Using these relations and replacing $\mathcal{W}$ and $\bar{\mathcal{W}}$ by $\mathcal{V}$ and $\mathcal{Z}$ in \eqref{eq:XiMagicFormula} one can deduce a suitable ansatz for $C$ and $\bar{C}$ in terms of $\mathcal{R}$ and $\mathcal{P}$ by demanding that up to $\mathcal{O}(\frac{1}{\ell^2})$ the l.h.s and the r.h.s of \eqref{eq:XiMagicFormula} have to agree. It turns out that a suitable ansatz for $C$ and $\bar{C}$ is given by
	\begin{equation}\label{eq:CBarCFlatRelations}
		C=\mathcal{R}+\frac{2}{\ell} D(\mathcal{R},\mathcal{P},\mathcal{M},\mathcal{N}),\quad \bar{C}=\mathcal{R}-\frac{2}{\ell}D(\mathcal{R},\mathcal{P},\mathcal{M},\mathcal{N}).
	\end{equation}
If one identifies
	\begin{equation}
		\frac{\mathcal{V}}{2\mathcal{M}^{\frac{3}{2}}}=\frac{\mathcal{R}-1}{\mathcal{R}^\frac{3}{2}},\quad \frac{\mathcal{Z}}{\mathcal{N}\sqrt{\mathcal{M}}}=\mathcal{P},
	\end{equation}
then $D(\mathcal{R},\mathcal{P},\mathcal{M},\mathcal{N})$ is given by
	\begin{equation}
		D(\mathcal{R},\mathcal{P},\mathcal{M},\mathcal{N})=\frac{\mathcal{N}}{\mathcal{M}}\frac{\mathcal{R} \left(\mathcal{R}^{\frac{3}{2}} \mathcal{P}+3 \mathcal{R}-3\right)}{ \left(\mathcal{R}-3\right)}.
	\end{equation}
Again, as in the spin-2 case the outer horizon limit does not yield the correct result as can be easily shown by replacing all AdS quantities in \eqref{eq:SouterBTZSpin3} by their flat space counterparts and taking the $\ell\rightarrow\infty$ limit. This yields the following expression
	\begin{equation}
		S_{\textrm{\tiny outer}} =\frac{\pi}{6}\ell c_M\sqrt{\mathcal{M}}\sqrt{1-\frac{3}{4\mathcal{R}}}+\mathcal{O}(\frac{1}{\ell}),
	\end{equation}
which is divergent with $\ell$ and thus, as expected, the outer horizon formula \eqref{eq:SouterBTZSpin3} is not the correct expression for a contraction to flat space. In close analogy to the spin-2 case I will assume that the following formula is an appropriate microstate counting of the \emph{inner} horizon entropy of a spin-3 charged BTZ
	\begin{equation}\label{eq:SinnerBTZSpin3}
		S_{\textrm{\tiny inner}} =2\pi\left|\sqrt{\frac{c\,\mathcal{L}}{6}}\sqrt{1-\frac{3}{4C}}-\sqrt{\frac{\bar{c}\,\bar{\mathcal{L}}}{6}}\sqrt{1-\frac{3}{4\bar{C}}}\right|,
	\end{equation}
whose only difference, as compared to the outer horizon formula, is again a relative minus sign between the two left- and right-moving contributions. In addition the $C\rightarrow\infty$ and $\bar{C}\rightarrow\infty$ limit yields the correct expression for the spin-2 inner horizon formula as it should be. Using \eqref{eq:SinnerBTZSpin3} and performing the same steps as before one obtains the following final result after taking the $\ell\rightarrow\infty$ limit
	\begin{equation}\label{eq:SGCASpin3Final}
		\boxed{S_{\textrm{GCA}}^{\textrm{Spin-3}}=\frac{\pi}{6}\left|c_L\sqrt{\mathcal{M}}\sqrt{1-\frac{3}{4\mathcal{R}}}+c_M\frac{\mathcal{N}\left(4\mathcal{R}-6+3\mathcal{P}\sqrt{\mathcal{R}}\right)}{4\sqrt{\mathcal{M}}(\mathcal{R}-3)\sqrt{1-\frac{3}{4\mathcal{R}}}}\right|.}
	\end{equation}
As expected the $\mathcal{R}\rightarrow\infty$ limit yields again \eqref{eq:cl4}. It is also important to note that the part of \eqref{eq:SGCASpin3Final} which is proportional to $c_M$ can alternatively also be derived by solving holonomy conditions (as in the AdS case) \cite{Gary:2014aa}. This shows that my assumption \eqref{eq:SinnerBTZSpin3} is not only plausible but seems indeed to be the correct expression for the inner horizon entropy of a spin-3 charged BTZ black hole as its \.In\"on\"u--Wigner contraction leads to the correct flat space result.\\

It would be interesting to generalize the results of this note to calculate the entropy of FSCs in flat space higher-spin gravity for general spin-N \cite{Gonzalez:2013oaa,Grumiller:2014lna} and to the flat limit of exotic BTZ black holes \cite{Townsend:2013ela}.
\vskip 1.6cm
\centerline{\bf Acknowledgements} \vskip 0.2cm \noindent
I am very grateful to Arjun Bagchi and Daniel Grumiller for proposing this problem to me and for the discussions we had. I would also like to express my thanks to Tadashi Takayanagi for discussions and the opportunity to pursue my research at the Yukawa Institute for Theoretical Physics at the University of Kyoto. MR is supported by the FWF project I 1030-N27.\\\\
\emph{Note added:} While finishing this note I was informed that Fareghbal and Naseh \cite{Fareghbal:2014qga} independently arrived at one of the main results \eqref{eq:cl4} through an \.In\"on\"u--Wigner contraction.

%\bibliographystyle{apsrev}
%\bibliography{review}

\end{document}